# High-Speed Serial Optical Link Test Bench Using FPGA with Embedded Transceivers

Annie C. Xiang [a], Tingting Cao [a], Datao Gong [a], Suen Hou [b], Chonghan Liu [a], Tiankuan Liu [a],
Da-Shung Su [b], Ping-Kun Teng [b], Jingbo Ye [a]

[a] Department of Physics, Southern Methodist University, Dallas, TX 75275, U.S.A
[b] Institute of Physics, Academia Sinica, Nangang 11529, Taipei, Taiwan

cxiang@smu.edu

*Abstract*

We develop a custom Bit Error Rate test bench based on Altera's Stratix II GX transceiver signal integrity development kit, demonstrate it on point-to-point serial optical link with data rate up to 5 Gbps, and compare it with commercial stand alone tester. The 8B/10B protocol is implemented and its effects studied.

A variable optical attenuator is inserted in the fibre loop to induce transmission degradation and to measure receiver sensitivity. We report comparable receiver sensitivity results using the FPGA based tester and commercial tester. The results of the FPGA also shows that there are more one-to-zero bit flips than zero-to-one bit flips at lower error rate. In 8B/10B coded transmission, there are more word errors than bit flips, and the total error rate is less than two times that of non-coded transmission. Total error rate measured complies with simulation results, according to the protocol setup.

## I. Introduction

High-speed serial optical data link provides a solution to High Energy Physics experiments' readout systems with high bandwidth, low power, low mass and small footprint. Many gigabits per second links are currently deployed at CERN's Large Hadron Collider (LHC) such as GLINK in calorimeter readout [1] and GOL in silicon tracker [2]. Next generation of multi-gigabit per second link is widely proposed to be operated at the Super LHC upgrade [3].

In the mean while, commercial FPGAs with embedded multi-gigabit transceivers have become readily accessible. Altera's Stratix II GX family and Xilinx's Virtex 5 FXT family offer comprehensive data interface designs that operate up to 6.5 Gbps. The newest Stratix IV GX and Virtex 6 HXT push the serial transceiver rate up to 10 Gbps.

These reconfigurable transceivers combined with programmable logic fabric make it feasible to develop a custom Bit Error Rate Tester (BERT) capable of verifying and characterizing a wide range of digital communication systems. Once the performance of the transceiver is verified, it can be deployed to demonstrate link architecture at system level. Compared with traditional standalone BERT equipment, FPGA-based BERT is much cheaper. It is also feasible to set up for different DUTs in irradiation tests due to its portable size and accessibility. Several groups have reported BER tests in a number of Single Event Effects (SEE) studies on optical and electrical components. [4][5]

Expedient customization is another major advantage of FPGA implementation. Reconfigurable hardware and build-in IPs support flexible prototyping. Function blocks are encapsulated and pluggable, for example, 8B/10B encoder and decoder can be enabled or by-passed to emulate different system architects. It is important to understand how these standard communication protocols affect the transmission of event data as well as time, trigger and control information. PC user interface through USB is also important to support real-time access of detailed error loggings for studying these effects as well as link degradation due to irradiation.

A set of Bit Error Rate tests are performed, which is also known as the receiver sensitivity tests. Using the same physical transmission link between transceivers, we report comparable results using FPGA based BERT and the commercial tester. We also conduct tests using non-coded data and 8B/10B encoded data, and compare the results to that of simulation.

## II. Test Bench Setup

### A. Optical link

We develop the test bench based on Altera's Stratix II GX transceiver signal integrity development kit and demonstrate it on a point-to-point serial optical link. A picture of the test bench set up is shown in Figure 1.

The FPGA-based BERT generates pseudo-random binary sequence (PRBS) at 5 Gbps. Its embedded transmitter drives a differential pair of coaxial cable that is connected to a SFP+ module. The SFP+ module consists of an optical transmitter (laser diode) and an optical receiver (photo diode). They convert the serial signal from electrical to optical and from optical back to electrical. The light output of the optical transmitter is coupled into two meters of OM3 grade multiple mode fibre. A variable attenuator is inserted in the fibre loop. The attenuator can be manually or automatically controlled. Fibre from the attenuator is plugged back into the optical receiver of the same SFP+ module. A carrier board is designed in house, onto which the SFP+ transceiver module is plugged. The board is impedance matched for high speed traces, and provides power and configuration to the module. Another pair of coaxial cables loops the electrical output signal of the optical receiver back to the FPGA's embedded receiver. The Altera development kit supports communication with a PC through USB port via FTDI interface. A user interface panel is coded in LabVIEW to download configurations and upload error loggings to and from the FPGA.

The physical media dependent portion of the data link begins and ends at the input and output coaxial cables,

inclusive. A stand alone commercial BERT is plugged in the place of the FPGA-based BERT for comparison. Data are collected on a set of SFP+ from various manufacturers and a set of different length fibre loops. There are no discrepancies among the test results and only results of one scenario are detailed in section 3.

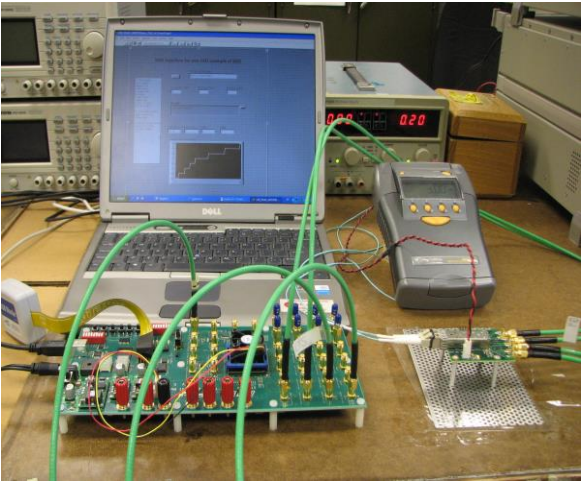

Figure 1: Setup for FPGA-based BERT driving serial optical link

### B. FPGA with embedded transceiver

The Stratix II GX FPGA dedicates the right side banks to transceiver circuitry that transmit and receive high-speed serial data streams. Each transceiver supports a number of protocols and operation modes with embedded hardware blocks and build-in firmware IPs [6].

We instantiate the transceivers to operate in basic mode through provided mega-function. And the instantiation is illustrated in Figure 2.

The FIFO buffer decouples clock phase variations across the programmable logic device (PLD) and the transceiver domains. Byte serializer allows the PLD to run at half the clock rate in order to match the transceiver speed. Byte ordering block is used in conjunction with byte deserializer to ensure the least and most significant byte order. Double-widths data path of the channel serializer and channel deserializer are enabled to support data rate of 5 Gpbs. Two cascaded 8B/10B encoders and decoders can be enabled or by-passed. The channel data path is 32 bits wide for non-coded transmission and 40 bits wide for 8B/10B transmission.

On board 156.25 MHz oscillator is enabled as the input reference clock for the transmitter and receiver clock synthesisers to generate required frequencies. The clock recovery unit works in automatic lock mode, i.e., it initially locks to the reference clock and then switches over to the incoming data stream.

Word aligner detects specific patterns, aligns word boundaries and flags link synchronization according to protocol specific or custom defined state machine. We use the same specified word pattern for alignment, ordering and synchronization for simplicity. It therefore may require several resets to achieve true synchronization, where all status flags are asserted.

Dynamic reconfiguration supports switching of analogue settings such as pre-emphasis, equalization and differential voltage amplitude at run-time through on-board dip switches and push buttons.

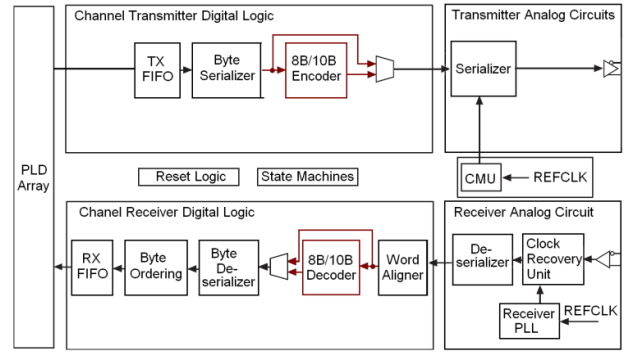

Figure 2: Simplified diagram of the transceiver implementation in hardware and firmware

### C. Pattern generator and error detector

Pattern generator and error detector are custom coded in the FPGA programmable logic fabric, in conjunction with the embedded transceiver, to generate and verify data stream that pass through the physical optical link. Pseudo-random binary sequence (PRBS) of length $2^7-1$ and $2^{23}-1$ are implemented in polynomial shifters as basic test patterns. Only results of $2^7-1$ PRBS are reported in section 3.

The functions of the pattern generator and error detector and their interfaces with the transceiver are controlled by state machines as illustrated in Figure 3.

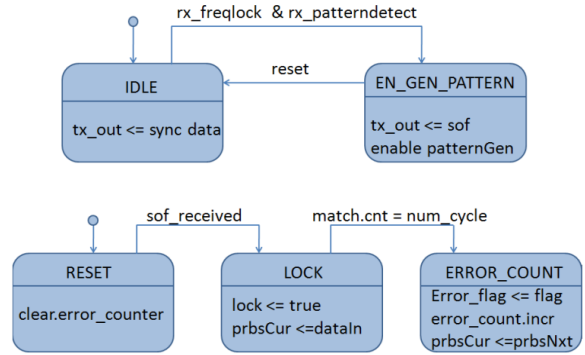

Figure 3: State machine of pattern generation (above) and error detection (below)

After power on or reset, synchronization data is sent from transmitter to receiver until frequency is locked and word alignment is achieved. Pattern generator is then enabled. The error detector uses the incoming data as seed to generated expected output data, until pattern match is declared. The error detector then switches to internal seed. Therefore, when the link is stable, incoming erroneous bit cannot disturb the output generation of error detector. Pattern match is declared when error-free incoming data is received for a specified number of consecutive clocks. Pattern match is not deserted, however, for consecutive error cycles. The frequency lock indicator will flag if the error cycles lead to link losing synchronization.

Error injection that simulates single bit flip is provided by XOR the least important bit. Error types, type counters and time stamps are logged in FIFOs for user access. Error statistics are performed on the PC side.

## III. RESULTS

### A. Signal integrity

We measure signals at several test points along the serial optical data link using oscilloscopes' electrical or optical modules. The test points and test parameters are illustrated in Figure 4. Test point 1 measures channel transmitter output. Test point 2 measures optical transmitter output. Test point 3 measures optical receiver input and channel receiver input is measured at test point 4. Example eye diagrams of the channel transmitter output and channel receiver input are shown in Figure 5. Zero pre-emphasis setting results in the best eye opening at the far end of optical receiver output. Transmitter PLL bandwidth, equalization and DC gain have no effect on the error rate performance under this test scenario.

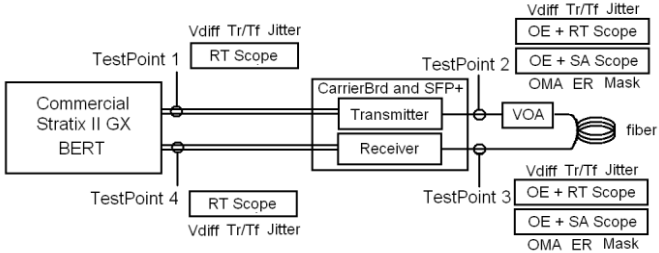

Figure 4: Schematic of test points and test parameters along a complete serial optical data link. (RT: real-time; SA: sampling; OE: optical-electrical converter; VOA: variable optical attenuator.)

The goal of measuring waveforms at different test points is to assign power (vertical) and jitter (horizontal) budget along these interfaces so that components that comply with these values would work together as one system. There are several industrial standards such as 4G Fibre Channel and 10GbE [7][8] that provide component acceptance value. For our application purpose, we must ensure that irradiation degradation is also accommodated while referring to these values. Jitter measured at channel transmitter output is 45ps, or 0.225 UI (unit interval), where the unit interval is 200ps for 5Gbps transmission. It is below both reference values from the 4GFC and 10GbE scaled. This validates the use of Stratix II GX transmitter to characterize downstream components. Jitter measured at channel receiver input is 60ps or 0.30 UI. This number is the convoluted contributions of channel transmitter, optical transceiver, and fibre loop. The difference of this value and the jitter acceptance value of the channel receiver is available for assignment to system degradations, such as fibre dispersion, connectors and irradiation. The rise/fall time of around 45 ps at channel transmitter output and rise/fall time of around 55 ps at channel receiver input also validate the use of the embedded transceiver to characterize link components and evaluate system bit error rate performance.

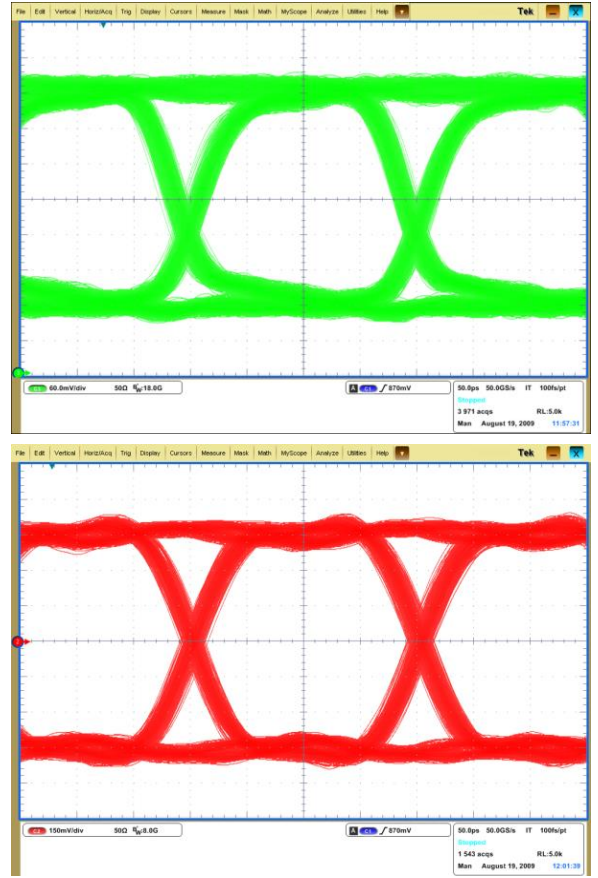

Figure 5: Eye diagram of the near-end transmitter (test point 1, above) and far-end receiver input (test point 4, below) of 5 Gbps, PRBS $2^7-1$ data pattern at room temperature.

### B. Basic BER

A variable optical attenuator is inserted in the fibre loop of the optical data link to induce transmission degradation. Bit error rate is measured at different attenuation levels. This test is also used to characterize the receiver sensitivity, the minimum optical power for achieving a specified bit error rate, i.e. at $10^{-12}$. In the noise dominate region, this relationship follow the general trend of error function of Gaussian distribution, where discrepancies are attributed to system nonlinearities as power penalties.

We compare the measurement results of the FPGA based BERT and that of a commercial BERT. The results are shown in Figure 6. The two testers obtain the same receiver sensitivity value for the same data link. The commercial BERT result deviates from the FPGA based BERT result as the bit error rate increases. This difference is due to the setup where the commercial BERT uses the same clock for both channel transmitter and receiver, whereas the FPGA based BERT has the ability to use clock recovered from the data stream as the channel receiver clock to mask out part of the system jitter.

We also observe that there is more one-to-zero bit flips than zero-to-one bit flips at lower error rate. This is due to the post amplification circuitry design of the optical receiver, which favourites one state over the other.

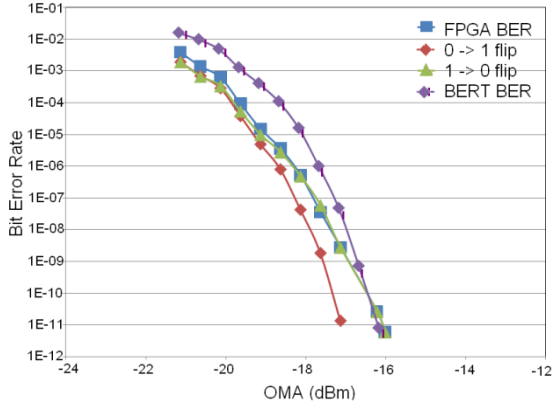

Figure 6: Bit error rate as a function of received optical power for 5Gbps, non-coded PRBS $2^7-1$ data transmission

### C. 8B/10B word error

The 8B/10B coding is used by many protocols to achieve: DC balanced data stream; sufficient level transitions; and unique code groups. Stratix II GX devices support two dedicated 8B/10B encoders in each transceiver channel. It works in cascade mode and complements the word aligner to achieve boundary synchronization.

The 8B/10B coding algorithm is implemented per 802.3ae standard [8]. In such a setup, a single bit flip in the serial data stream can affect one code group, resulting in multiple bit errors; or affect two code groups, resulting in invalid codes. When an affected code group is diagnosed as invalid, the output of the decoder is irrelevant. It is therefore simpler to record word error instead of bit error in this case. When the single bit flip induced error spread into multiple code groups, the propagation delay is uncertain, depending on the transmitted data. Error propagation is eventually stopped by nonzero disparity blocks and the timing distribution of propagation delay decreases rapidly. This knowledge is important to building criteria for evaluating coding schemes that can potentially cause inter-event interference such as in the case discussed above.

Figure 7 shows the Monte-Carlo simulation results of the error position distribution of 8B/10B coded transmission when the non-coded transmission bit error rate is $10^{-4}$. A total of 10,000 errors are inflicted, which is equivalent to $10^8$ bits in the serial data stream.

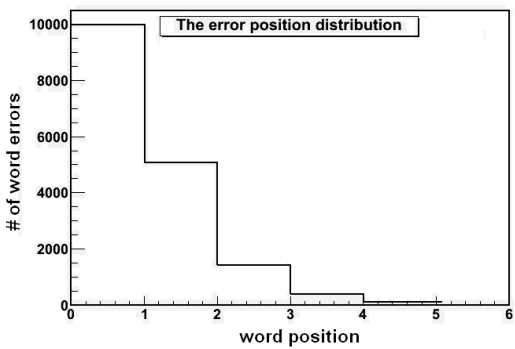

Figure 7: Simulation result of error rate of 8B/10B encoded data transmission when the non-coded serial data error rate is $10^{-4}$.

Table 1 shows the results of the simulation repeated at several levels of error rates. The majority of errors are word errors resulted from invalid codes. Most word errors occur in the first word after the bit flip (50%) as compared to the same word (18%) of the bit flip, and much less errors occur in the second word and insignificant amount occurs thereafter. Bit errors are also restricted to the same word of the bit flip.

Table 1: Monte-Carlo results of error rates of 8B/10B coded transmission with different non-coded transmission error rates

| serial err. rate | $10^{-4}$ | $10^{-6}$ | $10^{-8}$ | $10^{-10}$ |
|---|---|---|---|---|
| # of err. injected | 9997 | 10000 | 10000 | 10000 |
| total bit flip err. | 7239 | 7068 | 7175 | 7197 |
| total word err. | 13469 | 13618 | 13526 | 13632 |
| word err. without spread | 1774 | 1743 | 1751 | 1673 |
| err. spread to 1st word | 5135 | 5125 | 5094 | 5134 |
| err. spread to 2nd word | 1393 | 1392 | 1420 | 1446 |
| err. spread to 3rd word & later | 532 | 568 | 558 | 586 |

Error rate measurements of both non-coded and 8B/10B coded transmission are performed using the FPGA-based BERT. The results are shown in Figure 8. It confirms that there are more word errors than bit errors. And that that total word errors of 8B/10B transmission is less than two times that of the non-coded transmission.

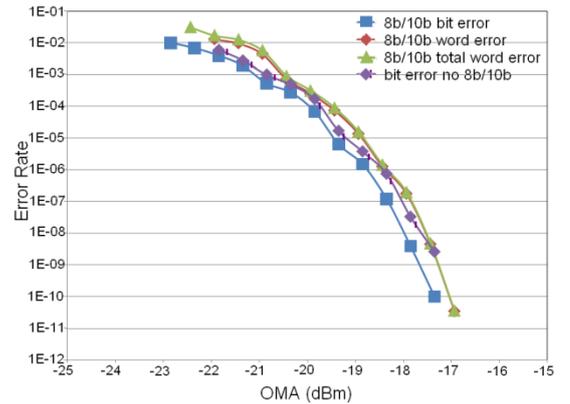

Figure 8: Bit errors and word errors as a function of received optical power for 5Gbps, non-coded PRBS $2^7-1$ data transmission vs. 8B/10B coded data transmission

### IV. CONCLUSIONS

A test bench of high-speed serial optical link using Altera's Stratix II GX transceiver SI development kit is demonstrated. Its performance satisfies the tentative requirements for 5Gbps point-to-point data link applications. Optical receiver sensitivity test results comply in between the FPGA setup and that of a standalone commercial BER Tester.

The development of a custom BER tester allows us to investigate detailed statistics of the errors. We report that there are more one to zero bit flip than zero to one bit flip at lower error rate due to the optical receiver circuitry deployed.

Word error rate and error propagation of 8B/10B protocol is analyzed and simulated. We implemented the 8B/10B coding block in the FPGA-BERT and the measurement results comply with simulation results. The timing distribution of error propagation will prove important in evaluating the coding scheme appropriate to event data acquisition in experiments adopting such links.

## V. Acknowledgements

The authors acknowledge US-ATLAS R&D program for the upgrade of the LHC, and the US Department of Energy grant DE-FG02-04ER41299. We would also like to acknowledge Drs. Francois Vasey, Jan Torska and Paschalis Vichoudis at CERN for beneficial discussions.